\documentclass[fleqn,usenatbib]{mnras}

\usepackage{newtxtext,newtxmath}
\usepackage[T1]{fontenc}
\usepackage{ae,aecompl}
\usepackage{graphicx}
\usepackage{amsmath}
\usepackage[normalem]{ulem}

\newcommand{\caps}[1]{{\scshape{#1}}}
\def\apj{ApJ}
\def\aj{AJ}
\def\mnras{MNRAS}

\def\aap{A\&A}
\def\apjs{ApJS}

\def\bb{SDSS1057\ }
\newcommand{\ms}{M$_{\sun}$}
\newcommand{\rs}{R$_{\sun}$}

\title[The period bouncer \bb]{The period bouncer system SDSS J105754.25+275947.5: first radial velocity study}

\author[Echevarr\'ia et al.]{
J. Echevarr\'ia,$^{1}$\thanks{e-mail:  jer@astro.unam.mx}
S. Zharikov,$^{2}$
and
I. Mora Zamora$^{1}$,
\\
$^{1}$Instituto de Astronom\'ia, Universidad Nacional Aut\'onoma de M\'exico, Ciudad Universitaria 04510, CDMX, M\'exico\\
$^{2}$Instituto de Astronomía, Universidad Nacional Autónoma de México, Apartado postal 106, C.P. 22800, Ensenada, B.C., México\\
}
\date{Accepted XXX. Received YYY; in original form ZZZ}

\pubyear{2020}

\begin{document}
\label{firstpage}
\pagerange{\pageref{firstpage}--\pageref{lastpage}}
\maketitle

\begin{abstract}
We report the first radial velocity spectroscopic study of the eclipsing period bouncer SDSS J105754.25+275947.5. Together with eclipse light curve modeling, we redetermined the system parameters and studied the accretion disk structure. 
We confirm that the system contains a white dwarf with $M_{\mathrm{WD}}=0.83(3)~ $\ms\ and an effective temperature of 11,500(400)~K. The mass of the secondary is $M_2=0.056~$\ms\ with an effective temperature of T$_2$~=~2,100~K or below. The system inclination is $i=84\fdg3(6)$. The data is in good agreement with our  determination of $K_1$ =  33(4) km\,s$^{-1}$. We estimate the mass transfer rate as $\dot{M}=$1.9(2)$\times 10^{-11}$\ms~yr~$^{-1}$.  Based on an analysis of the SDSS and  OSIRIS spectra, we conclude that the optical continuum is formed predominantly by the radiation from the white dwarf. The contribution of the accretion disk is low and originates from the outer part of the disk. The Balmer emission lines are formed in a plasma with $\log$~$N_0$ = 12.7 [cm$^{-1}$] and a kinetic temperature of T$\sim$~10,000~K. The size of the disk, where the emission lines are formed, expands up to $R_\mathrm{d,out}=0.29$~\rs.  The inner part of the emission line forming region goes down to $R_\mathrm{d,in}\approx 2 R_\mathrm{WD}$ . The Doppler tomography and trailed spectra shows the presence of a hot spot and a clumpy structure in the disk, with variable intensity along the disk position angle. There is an extended region at the side opposite the hot spot with two bright clumps caused more probably by non-Keplerian motion there.
\end{abstract}


\section{Introduction}
\label{intro}
Cataclysmic Variables (CVs) are interacting binaries, which are composed of a white dwarf (WD) primary star and a late--type spectral secondary,  which transfers matter to the usually more massive primary star. An accretion disk is formed around the primary star, and the material leaving the inner Lagrangian point collides somewhere along the already formed disk, producing a bright spot. \citet{wan71} and \citet{sma71} establish this classical model, which works rather well for most dwarf novae, nova-like variables, and old novae. 

SDSS J105754.25+275947.5 (hereinafter \bb) is an eclipsing  short orbital period $P_{\mathrm{orb}}=0.062807(43)$~d system which shows all the characteristics of a bona-fide bounce back system. It is generally assumed that CVs evolve toward shorter orbital periods reaching the period minimum, at which point the secondary star achieves a high level of electron degeneracy (becoming a brown dwarf) and further contraction of the orbit becomes impossible \citep{1981ApJ...248L..27P}. From that point the orbital period starts to increase and CVs form so-called post-period minimum systems frequently named as period bouncer systems. \bb was identified as a CV by \citet{sea09}. \citet{sou15} found that the white dwarf component mainly dominates its spectrum while the emission components of the Balmer lines are double-peaked. They also proposed \bb as a good candidate for a period-bounce system. \citet{mca17} presented high-speed, multicolour photometry. Using the fitting of a parametrized eclipse model of average light curves, they report the following: a mass ratio of $q$~=~$0.0546(20)$ with an inclination $i$~=~85\fdg74(21). The white dwarf and donor masses were found to be $M_{\mathrm{WD}}$=0.800(15)~\ms\ and $M_2$=0.0436(20)~\ms, respectively. The temperature was estimated as $T_{\mathrm{WD}} = 13,300(1,100)$ K and using a distance to the system of $d = 367(26)$ pc. The last is   larger than the new Gaia photo-geometric distances 302$_{-30}^{+12}$ pc obtained from EDR3 \citep{bai21}. \citet{mca17} also estimated a mass transfer rate in the system as $6.0_{-2.1}^{+2.9}\times10^{-11}$~\ms~yr$^{-1}$. The interstellar absorption is very low $E(g-r)=0.01_{-0.01}^{+0.02}$ in this direction at the source distance following \citet{green18}.

\begin{figure*}
\setlength{\unitlength}{1mm}
 \begin{center}
 \begin{picture}(170,170)(0,0)
\put (0,50) {    \includegraphics[width=17cm,bb = 10 20 750 590, clip=]{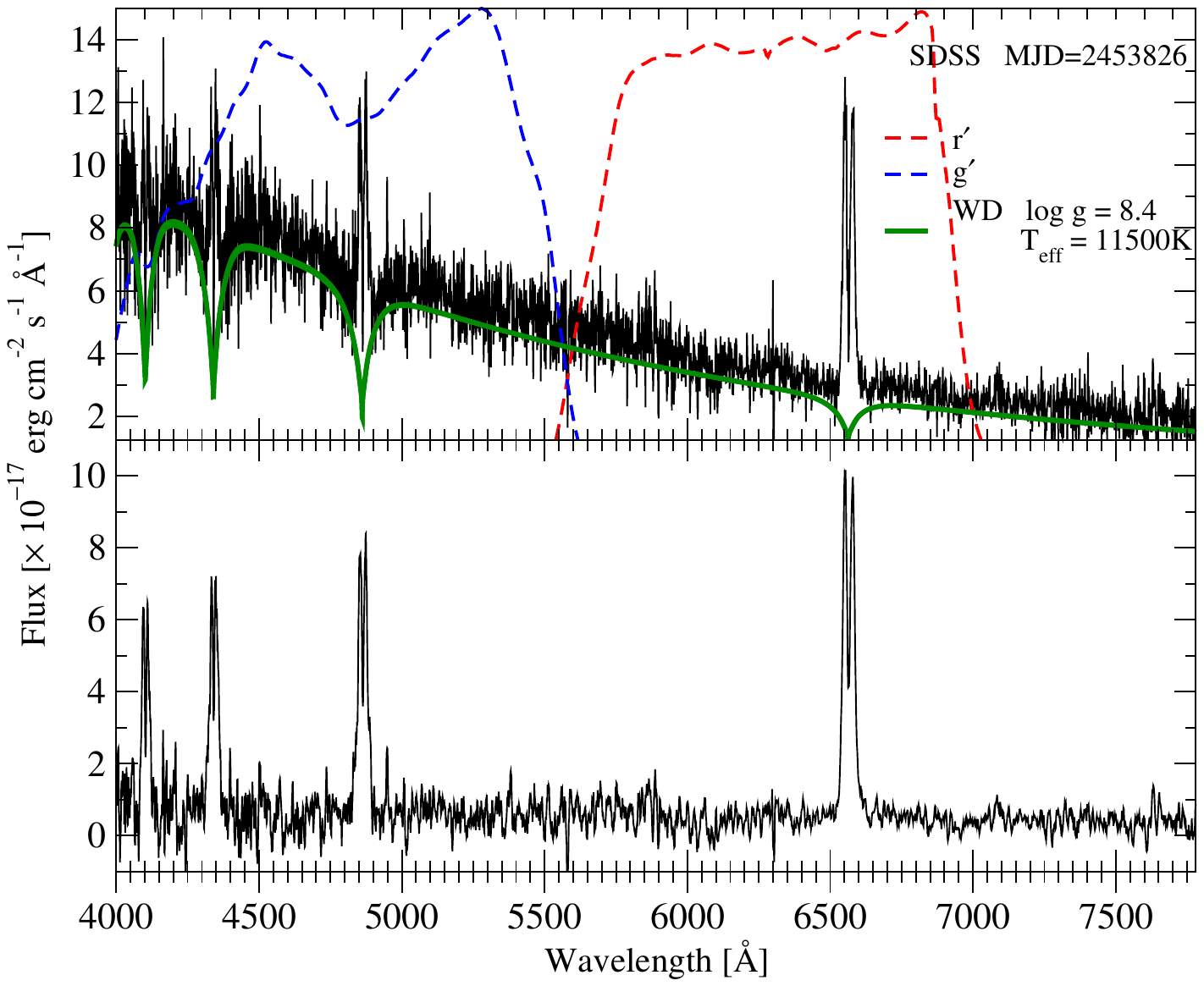}}
\put (-3,0) {\includegraphics[width=5.8cm,bb = 10 0 752 573, clip=]{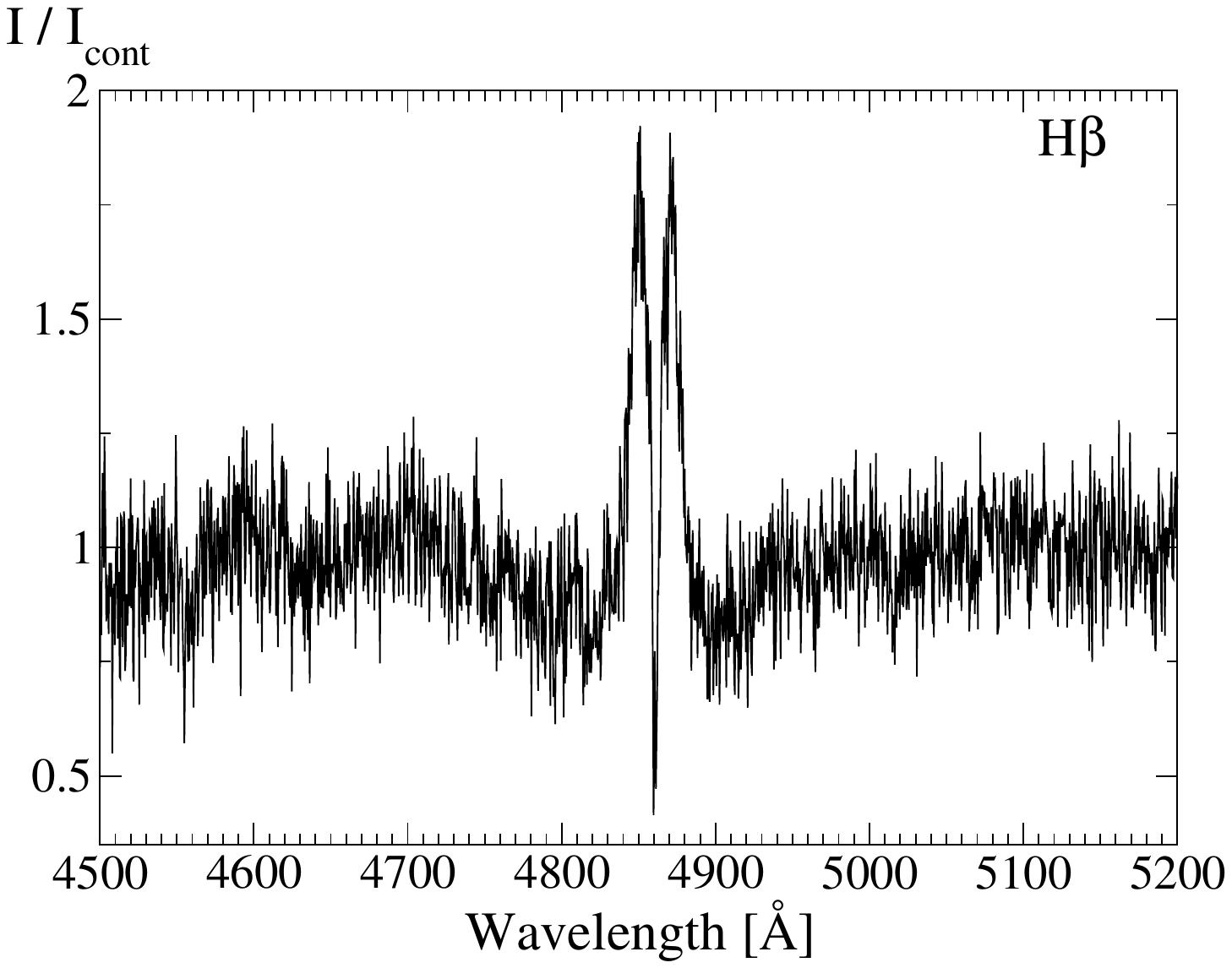}}
\put (57,0) {    \includegraphics[width=5.8cm,bb = 10 0 752 573,clip=]{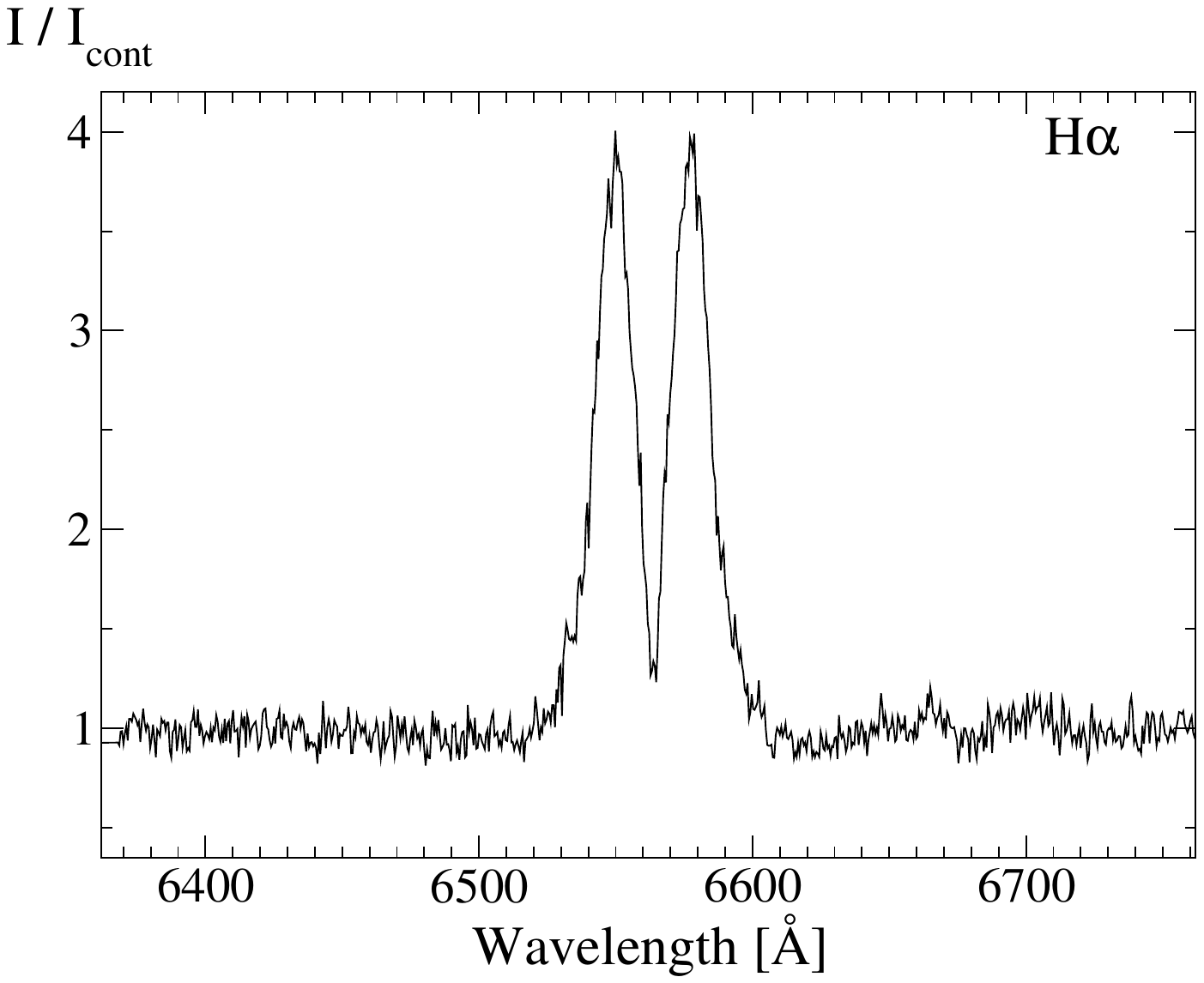}}
\put (118,0) {    \includegraphics[width=5.8cm,bb = 10 0 750 570, clip=]{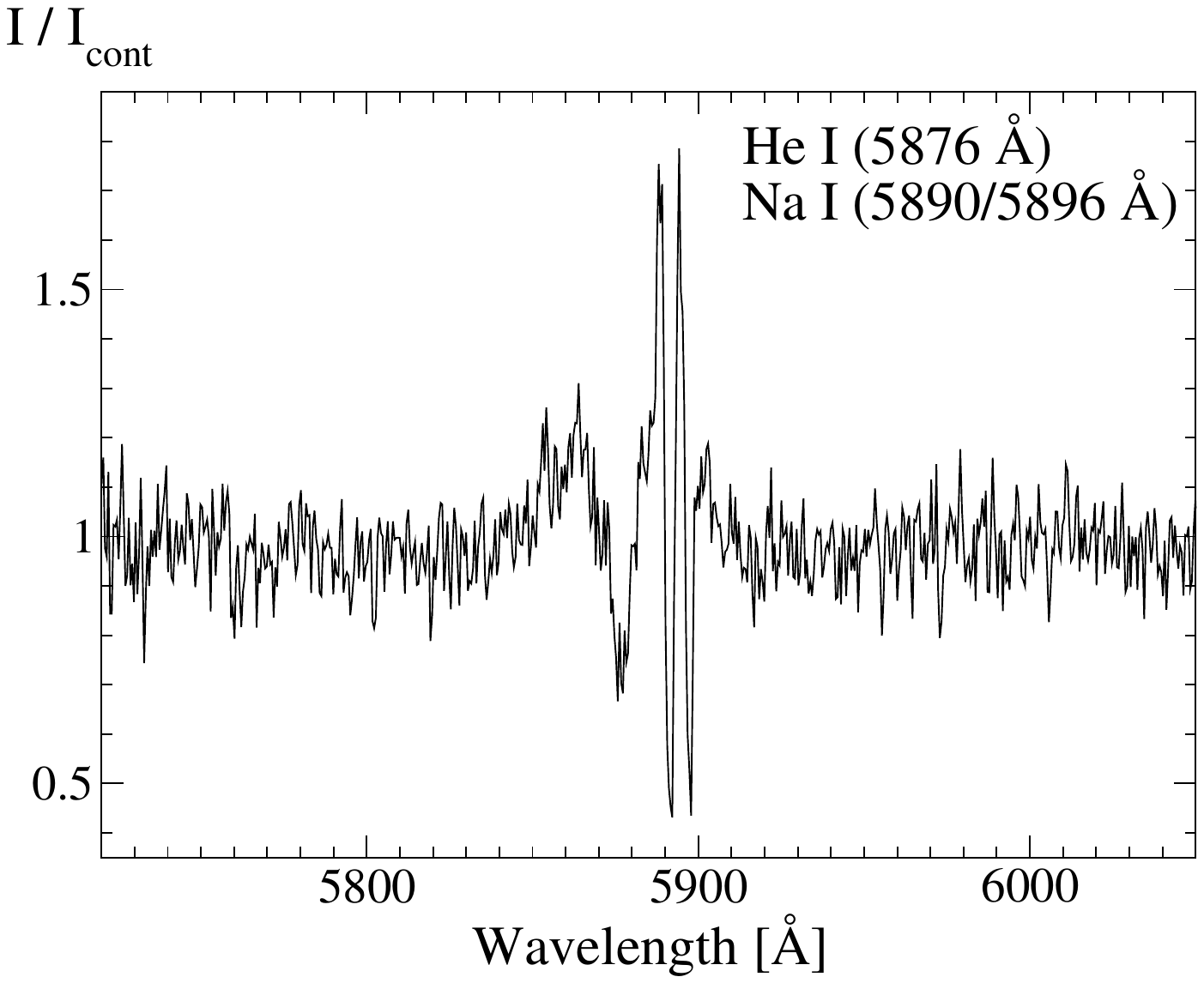}}
\end{picture}
\end{center}
\caption{Top panel: The SDSS spectrum of SDSS1057 (black solid line), the WD underlying spectrum (\citet{2010MmSAI..81..921K}, green solid line) with parameters obtained from light curve fitting  (see text below). Blue and red dashed lines mark the transparency curve of the g$^\prime$ and the r$^\prime$- ULTRACAM filters, respectively.
Middle panel: The accretion disk spectrum which is a result of  subtraction of the WD spectrum from SDSS one and smoothed by a running average of 7\AA.
Lower panels: The phase-average profiles of H$_\beta$ and H$_\alpha$ lines as well as the complex HeI-NaI lines in the OSIRIS spectrum of SDSS1057 normalized to continuum.
}
    \label{fig:HaHb}
\end{figure*}

In this paper, we report the first radial velocity study of \bb. In Section~\ref{sec:obs} we describe the spectroscopic observations. In Sections~\ref{sec:spec} and \ref{sec:radvel} we discuss the object spectrum and the radial velocities of H$\alpha$ and H$\beta$ emission lines.  In Section~\ref{lcm} we  demonstrate the result of a re-determination of the system parameters using recently obtained ULTRACAM photometry and our modelling tool, and in  Section~\ref{disc} we discuss the disk emission forming regions. In Section~\ref{tomo} we use Doppler tomography to help us understand the nature of the accretion disc,  and finally  in Section~\ref{sec:Concl}  our conclusions are given.

\section{Spectroscopic Observations}
\label{sec:obs}
Spectra were obtained with the 10.4~m telescope of the Observatorio Roque de los Muchachos (Gran Telescopio Canario) using the OSIRIS instrument\footnote{http://www.gtc.iac.es/instruments/osiris/} in the long-slit mode on the nights of 2023 February 19 and March 1. The log of observations of SDSS1057 is given in Table~\ref{tab:speclog}. On the first night, we used the grism R2500V (4500-6000\AA), and on the second   R2500R (5575-7685\AA), which gave a R$\sim$ 2500 resolution. In each observational run, we obtained 24 spectra with an individual exposure time of 235~sec which allowed us to cover about 94 min, a little more than the orbital period of 90.4 min.  Standard~{\sc iraf} procedures were used to reduce the data. 

\begin{table}
\centering
	\caption{Log of spectroscopic observations of \bb.}
    \label{tab:speclog}
    \begin{tabular}{lcccc}
        \hline
        Date & Julian Date & grism & esposure& No of    \\
        (dd/mm/yyyy)       & (2400000+) & & (sec)  &  spectra   \\
        \hline
        19/02/2023 & 59994   & 2500V & 235 & 24 \\
        01/03/2023 & 60004  & 2500R & 235 & 24 \\
       \hline
\end{tabular}
\end{table}       

\section{Optical spectrum}
\label{sec:spec}
\begin{figure}
\setlength{\unitlength}{1mm}
 \begin{center}
 \begin{picture}(80,66)(0,0)

 \put (-6,0) { \includegraphics[width=0.50\columnwidth, bb = 0 0 380 600, clip]{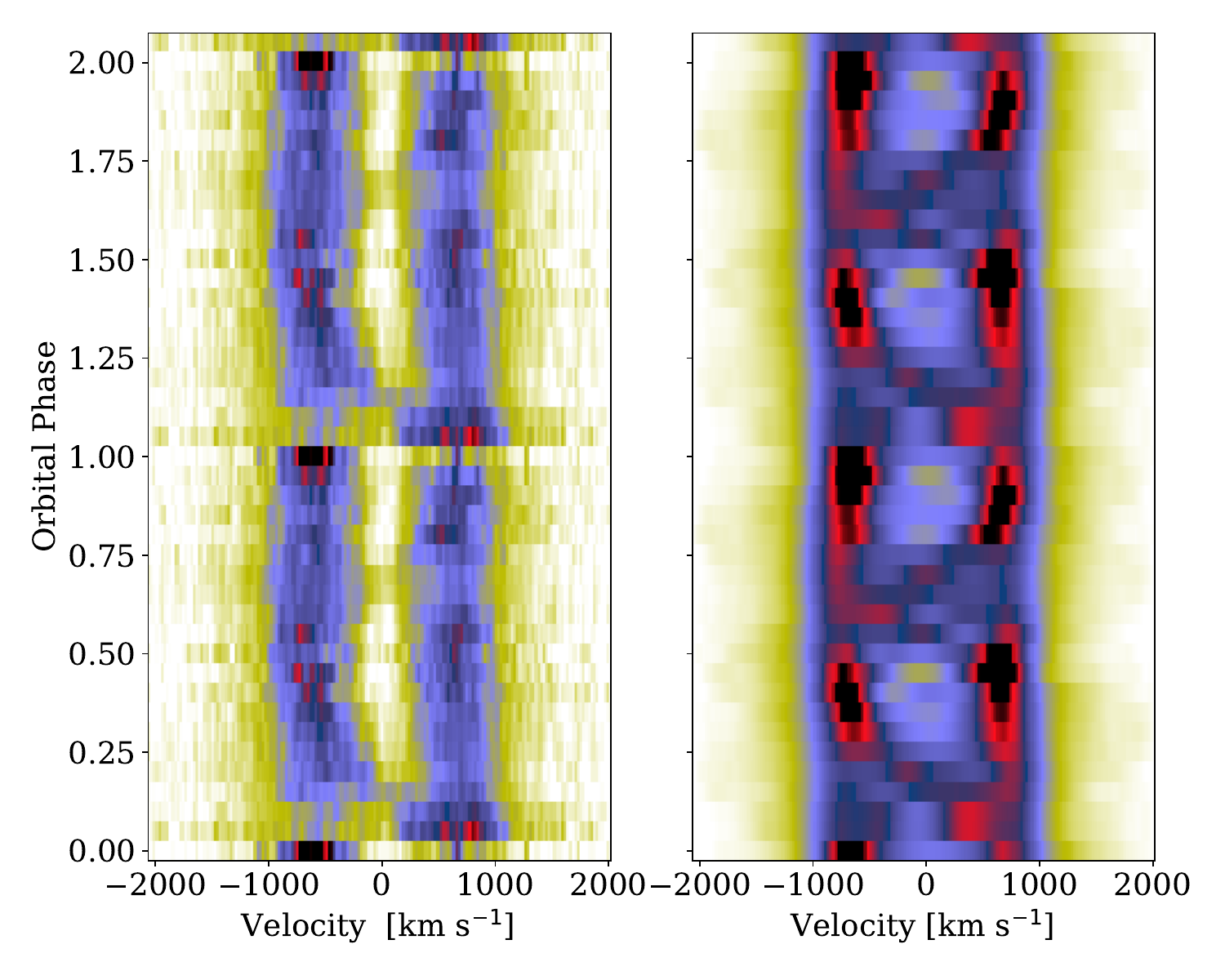} }
 \put (37,0)  { \includegraphics[width=0.52\columnwidth,  bb = 0 0 393 560, clip]{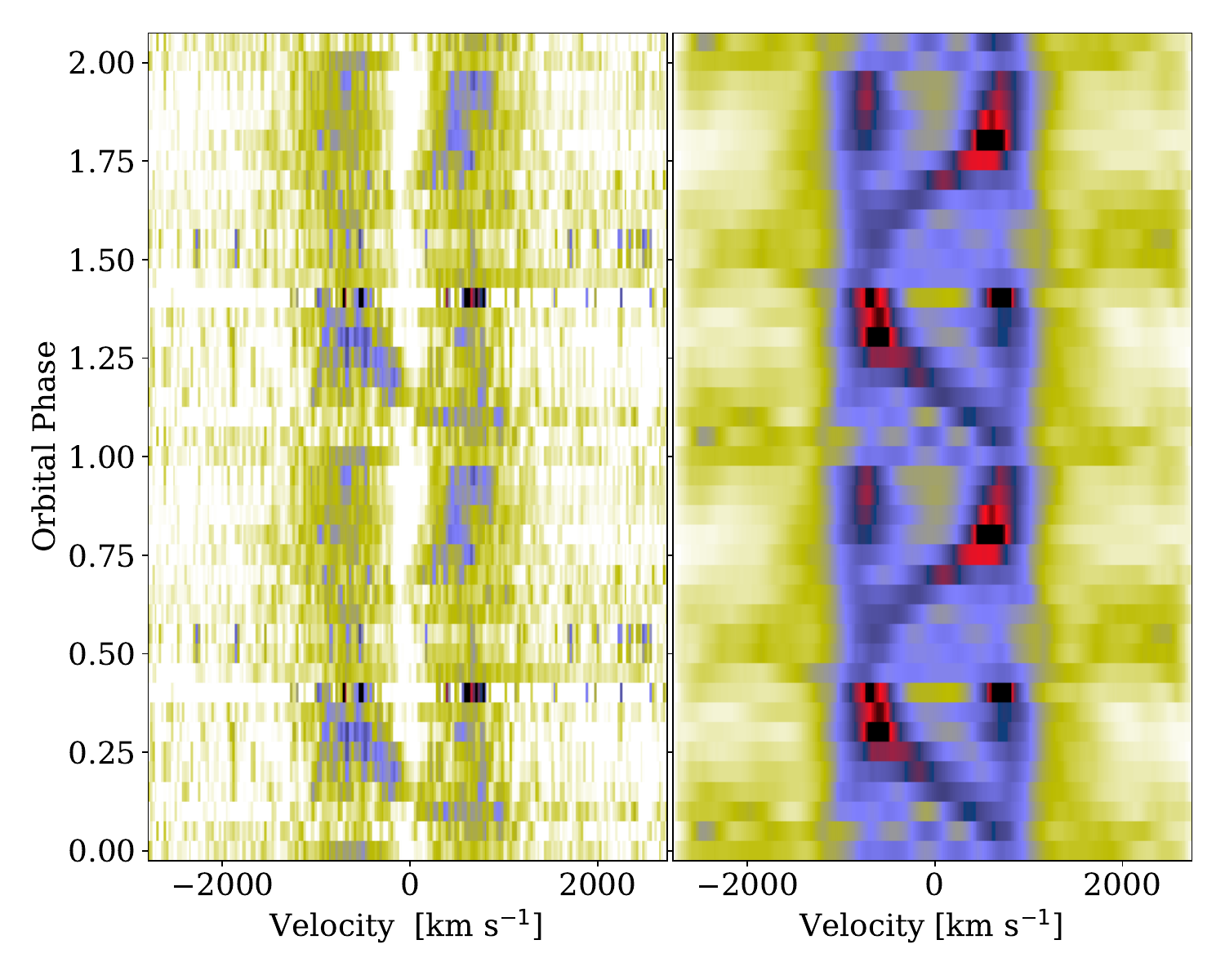}}
\put (5,59) {\it \Large H$\alpha$}
\put (50,59) {\it \Large H$\beta$}
    \end{picture}
\end{center}
    \caption{Observed trail of the H$\alpha$ (left) and H$\beta$ (right) emission lines. The relative emission intensity is shown in a scale of colors, where the strongest intensity is represented by black, followed by red, then blue, and finally yellow which depicts the weakest intensity. The color white represents the lack of emission.}
    \label{fig:TRSpec}
\end{figure}

A simple sum of GTC/OSIRIS spectra  without radial velocity  correction is shown for  H$\beta$, H$\alpha$, and \ion{He}{I} (5876\AA) in Fig.~\ref{fig:HaHb} (lower panels). For comparison, in the top panel of Fig.~\ref{fig:HaHb}, we show the SDSS flux calibrated spectrum of the source \citep{sea09}. The  lines in the OSIRIS spectra look  strong and appear double-peaked. The H$\alpha$ line is twice as strong as the H$\beta$ lines.  The latter shows clearly a broad absorption component arising from the white dwarf and also a central narrow absorption that goes below the continuum. The \ion{He}{I} (5876\AA) is also  detected and looks double-peaked, though the red peak is contaminated \ion{Na}{I} doublet. The trailed spectra of H$\alpha$ and H$\beta$ lines show a slight variation of intensity along the orbital phase, the presence of a weak S-wave from the hot spot, and a  jump of intensity and velocity before and after very closely  at the eclipse (see Fig.~\ref{fig:TRSpec}). The hot spot contribution to the total line intensity is relatively low and  varies strongly with the orbital phase. It is more intense after the eclipse at orbital phase of 0.15-0.30; decreases its brightness  between 0.3-0.6, and then after comes back to the previous flux level.

\begin{figure}
	\includegraphics[width=1.0\columnwidth, bb= 1 40 710 600, clip=]{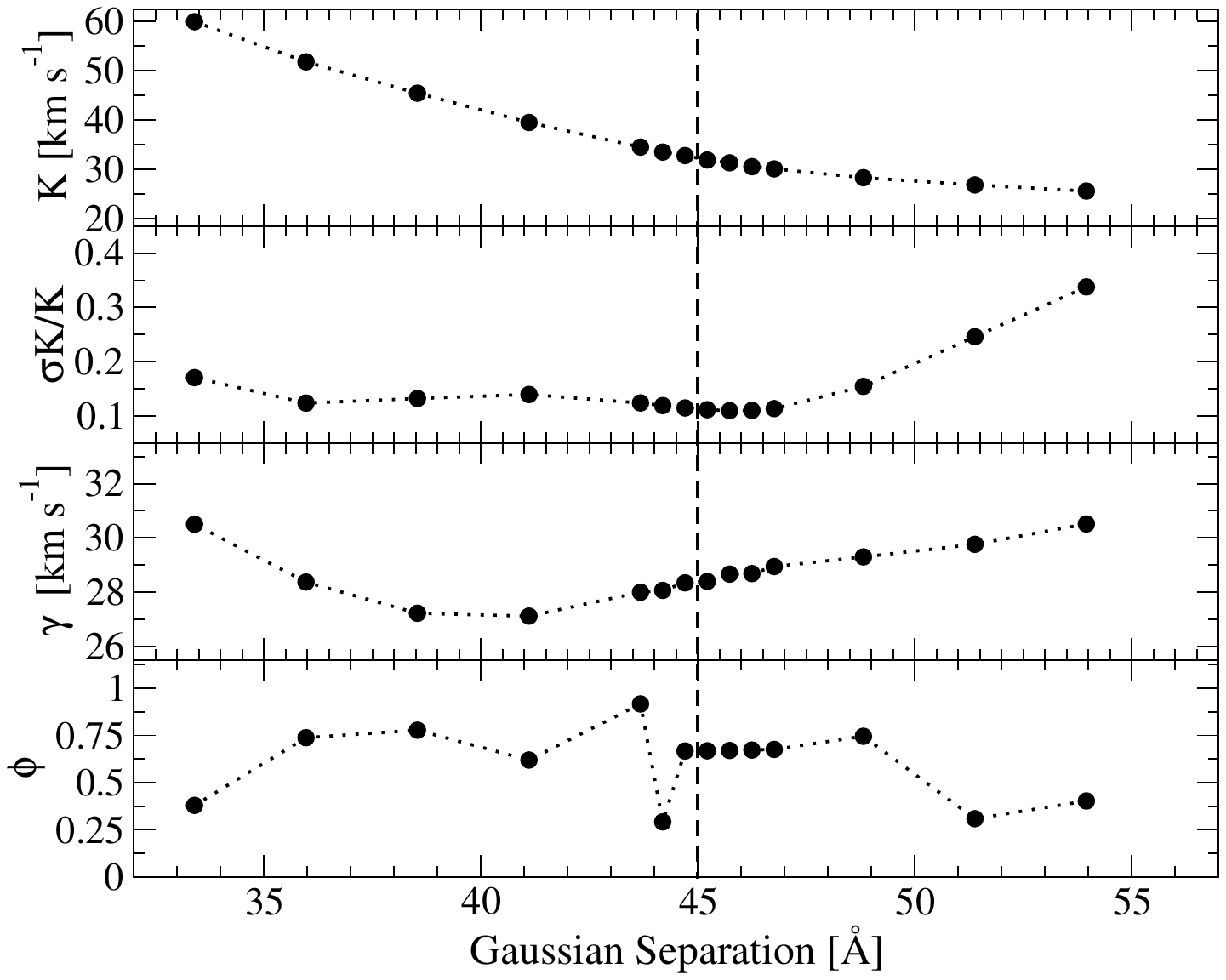}
    \caption{ Diagnostic Diagram of the radial velocity curve using the wings of the H$\alpha$ emission line. The best solution is for a Gaussian separation of about 45\AA\ and a width of about 5 \AA.}
    \label{fig:diag}
\end{figure}

\section{Radial velocity analysis}
\label{sec:radvel}

\subsection{The primary star}
\label{radvel1}

To measure the radial velocity of the emission lines, such that they reflect more accurately the motion of the primary star, \citet{say80} proposed a method to determine a more precise centre of the line by measuring the wings only; since they originate in the vicinity of the white dwarf, asymmetric-low-velocity features are avoided. This method, which uses two Gaussian functions with a fixed width and variable separation, was further expanded by \citet{sea86}, who developed a diagnostic diagram. In particular, they defined a control parameter, $\sigma_{K} / K$, whose minimum is a very good indicator of the best $K$ value. To measure the wings of the emission lines, we used a {\sc convrv} task within the {\sc rvsao} package in {\sc iraf}. This {\sc convrv} routine was given to us  by J. Thorstensen in 2008, who describes it as a task that computes the velocity of a line for a set of spectra by convolving the line with an antisymmetric function and taking the line center to be the zero of this convolution \citep[see][for full details]{eea21}. To produce a diagnostic diagram,    we have to iterate between using {\sc convrv}
and running the results using a simple orbital solution of the type:

\begin{equation}
V(t) = \gamma + K \sin\left(2\pi\frac{t - t_0}{P_{orb}}\right),
\end{equation}

where $V(t)$ are the measured radial velocities of the individual spectra, $\gamma$ is the systemic velocity, $K$ is the radial velocity semi-amplitude, $t_0$ is the time of inferior conjunction of the donor, and $P_\mathrm{orb}$ is the orbital period. The latter is well known from eclipse observations: 90.44 min \citep[see][]{mca17}. We set this orbital period as a fixed parameter 
and derived $K$, $\gamma$ and $t_0$ as free parameters. We used orbital\footnote{running \caps{orbital}which uses the
least square method to determine system orbital parameters. Available at \url{https://github.com/Alymantara/orbital_fit}}  with the parameters given in each run of {\sc convrv} for different Gaussians separations to finally produce the diagnostic diagram shown in  Fig.~\ref{fig:diag}. The best-fitted values were obtained for Gaussian separations of   45 \AA ~ with individual widths of~5 \AA.

\begin{figure*}
\setlength{\unitlength}{1mm}
 \begin{center}
 \begin{picture}(170,55)(0,0)
 \put (-2,0) { \includegraphics[width=1.0\columnwidth, bb = 0 40 710 540, clip]{{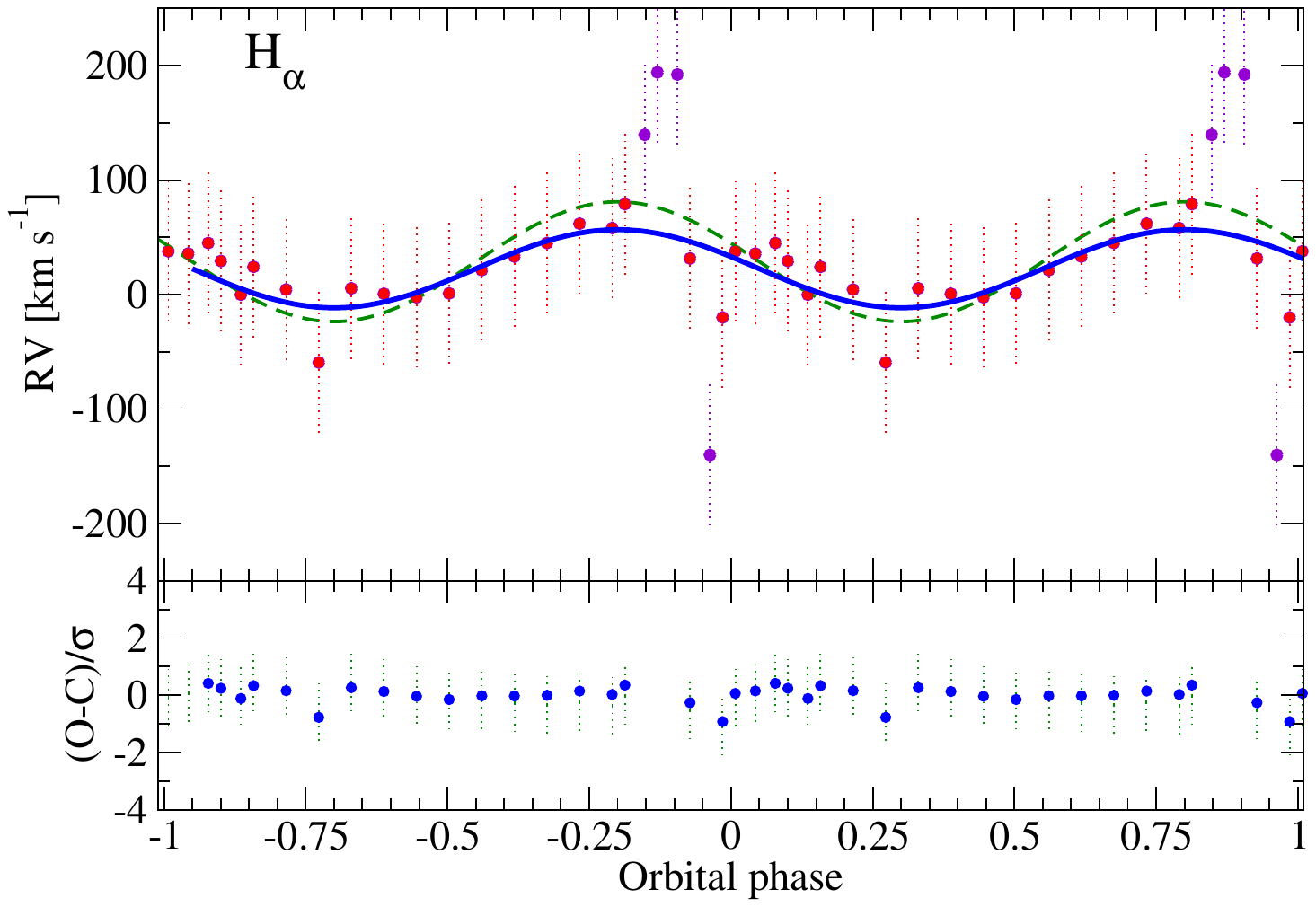}} }
 \put (85,0)  { \includegraphics[width=1.0\columnwidth,  bb = 0 40 710 540, clip]{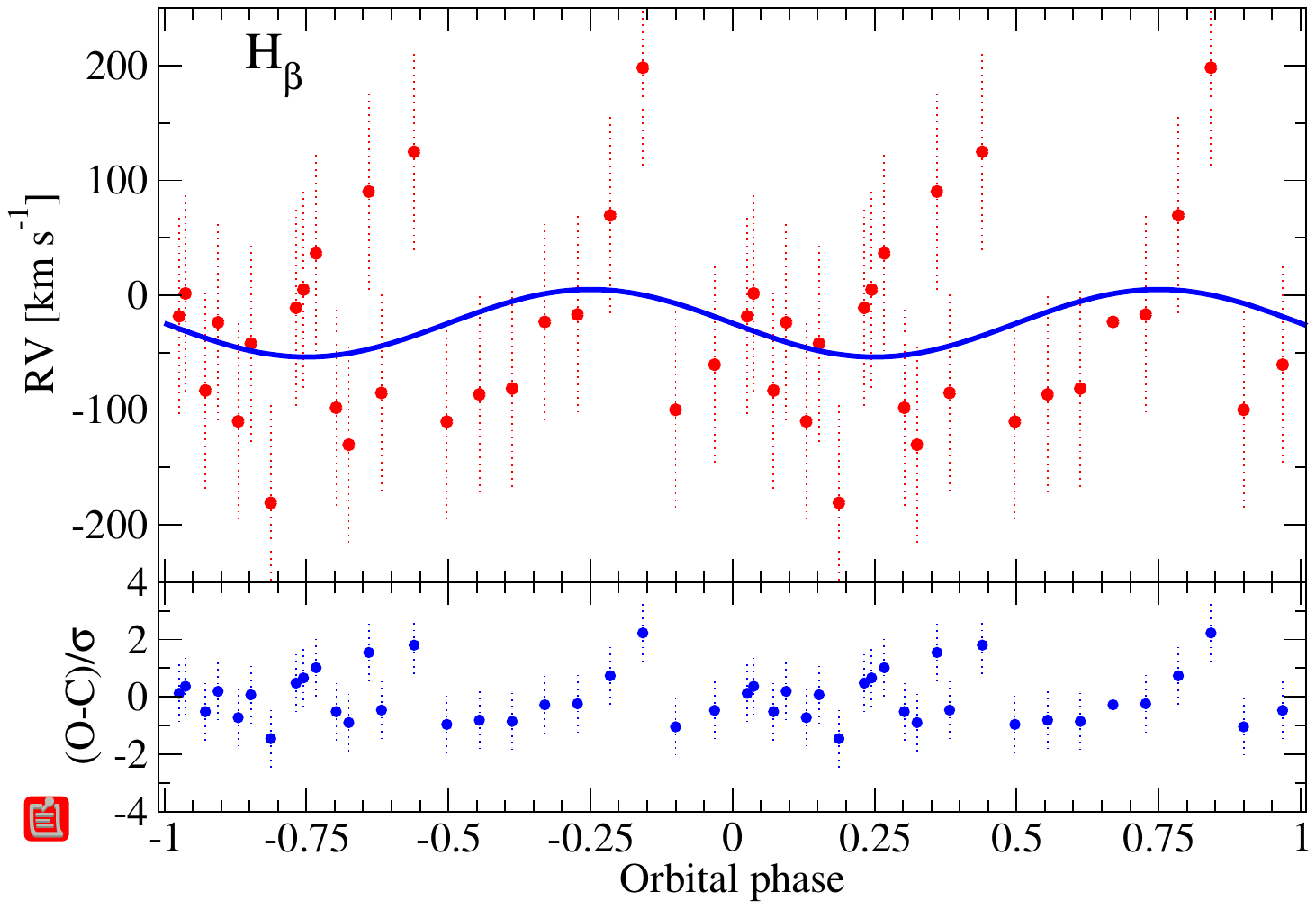}}
    \end{picture}
\end{center}

    \caption{Radial velocity curve of the wings of the H$\alpha$ (left)  and  H$\beta$ (right) emission lines. 
    For H$\alpha$ two solutions are shown. The first marked by  the green dashed line and includes the distorted velocities (the violet points) close to the eclipse, the second (blue solid line) is the solution for the measurements marked only by the red points. 
    The lower panes shows the (O-C)/$\sigma$ diagrams of the fits. }
    \label{fig:radial-distorted}
\end{figure*}
 
\subsection{Radial velocity curves}

The  result of the fit of the non-distorted points (marked by red in Fig.~\ref{fig:radial-distorted},  left) are shown in Table~\ref{orbpar}  (H$\alpha$ wings). The inclusion of the distorted points gives, of course, a larger and noisier semi-amplitude of $K_1$~=~35(12)~km\,s$^{-1}$ (the green dashed line in Fig.~\ref{fig:radial-distorted},~left). We have also attempted to fit the H$\beta$ spectra, but as they are much fainter than H$\alpha$, we have not obtained a good fit as shown in Fig.~\ref{fig:radial-distorted},~right. The sinusoidal blue line has a semi-amplitude of $K_1$~=~30(25)~km\,s$^{-1}$.

As mentioned by \citet{sea86}, it is instructive to plot $K_1$, $\gamma$, and $t_0$ (here shown as $\phi=t_0(a) - t_o$), as part of the diagnostic diagrams, since we expect them to behave in a particular way when approaching a good fit to the wings of the lines. These values, as well as $\sigma_\mathrm{K_1} / K_1$ should show a slow approach towards the optimal values, and then, as the separation increases, they should depart rapidly as the measurements are dominated by noise. i.e., we start to measure outside the wings, which is what we see in Fig.~\ref{fig:diag}.

\begin{table}
\centering
\caption{Orbital Parameters obtained from the H$\alpha$ line. } 
\label{orbpar}
\begin{tabular}{lc}
\hline
Parameter  &  H$\alpha$            \\
\hline
  $\gamma$ & 28(2) km\,s$^{-1}$  \\
   $K_1$ &  33(4) km\,s$^{-1}$  \\
  HJD$_0$*          & 0.88397(90) \\
  P$_{\mathrm{orb}}$  & Fixed** \\
\hline \\
\end{tabular}

\begin{tabular}{l}
$^*$(245993+ days)\\
$^{**}$ 0.0627919557 days (fixed)\\
\end{tabular}
\end{table}

\begin{table}
\centering
\caption{Parameters of eclipse light curves fit.}
\label{tab:DiskPar}
\begin{tabular}{lllccc}
\hline\hline
\multicolumn{6}{c}{\bf Fixed parameters}    \\  
\hline
\multicolumn{3}{l}{$P_{\mathrm{orb}}$}  &  \multicolumn{3}{c} {5426.52 s}   \\ 
\multicolumn{3}{l}{$E(g^\prime-r^\prime)$ }           &  \multicolumn{3}{c} {0.01}    \\
\multicolumn{3}{l}{Distance [pc]}       &  \multicolumn{3}{c} {302.0}  \\
\hline
\multicolumn{6}{c}{{\bf Variable and their best values}} \\  
\hline
\multicolumn{3}{l}{{\bf System parameters}} & & & \\ 
\hline
\multicolumn{3}{l}{ $i$ [degree]}                 & \multicolumn{3}{c}{84\fdg3(6) }  \\
\multicolumn{3}{l}{ $M_{\mathrm{WD}}$ [\ms]}      & \multicolumn{3}{c}{0.83(3) }  \\
\multicolumn{3}{l}{ ${T}_{\mathrm{WD}}$ [K] }     & \multicolumn{3}{c}{11500(400)  } \\ 
\multicolumn{3}{l}{ $T_{2}$ [K] }                 & \multicolumn{3}{c}{2100$^{+400}_{-1000}$}   \\
\multicolumn{3}{l}{ $q = M_2/M_{\mathrm{WD}} $}        & \multicolumn{3}{c}{ 0.068$^{+0.04}_{-0.06}$}  \\
\multicolumn{3}{l}{ $\dot{M}$ [$10^{-11}$ \ms~yr$^{-1}$]} &  \multicolumn{3}{c}{1.9(2)} \\
\hline
\multicolumn{4}{l}{{\bf Parameters of the disk}} \\ 
\hline\noalign{\smallskip}
\multicolumn{3}{l}{ $R_{\mathrm{d, in}}$ [\rs] }   &\multicolumn{3}{c}{0.147 }  \\
\multicolumn{3}{l}{ $R_{\mathrm{d, out}}$ [\rs]}  & \multicolumn{3}{c}{ 0.287 } \\
\multicolumn{3}{l}{ $z_{\mathrm{d, out}}$ [\rs]}   & \multicolumn{3}{c}{0.03 }\\
\multicolumn{3}{l}{ $\gamma_{\mathrm{disk}}$ (fixed) }  &   \multicolumn{3}{c}{1.125 } \\
\multicolumn{3}{l}{ $EXP$ (fixed) }   &\multicolumn{3}{c}{0.25 }  \\
\hline 
\multicolumn{3}{l}{{\bf The hot spot/line}}&   \\ \hline\noalign{\smallskip}
\multicolumn{3}{l}{length spot ($\varphi_{\mathrm{min}}+ \varphi_{\mathrm{max}}$) }  &\multicolumn{3}{c} {70\fdg0}  \\
\multicolumn{3}{l}{width spot (\%$R_{out}$) }  &\multicolumn{3}{c} {24}  \\
\multicolumn{3}{l}{Temp. excess spot ($T_{\mathrm{s, max}}/T_{\mathrm{d,out}}$)  }  &\multicolumn{3}{c} {2.04}\\
\multicolumn{3}{l}{Spot Z slope (fixed) }  &\multicolumn{3}{c} {-3}  \\
\multicolumn{3}{l}{Spot Temp. slope (fixed)  }  &\multicolumn{3}{c} {$-3$}  \\
\multicolumn{3}{l}{Z excess spot ($z_{\mathrm{s, max}}/z_{\mathrm{d,out}}$) (fixed) }  &\multicolumn{3}{c} {1.03}   \\
\hline
\multicolumn{4}{l}{{\bf Calculated}} \\ 
\hline
\multicolumn{3}{l}{$a$ [\rs]}       &\multicolumn{3}{c}{0.64  }  \\
\multicolumn{3}{l}{$M_{2}$ [\ms] }  & \multicolumn{3}{c}{ 0.056 } \\
\multicolumn{3}{l}{$K_{2}$ [km s$^{-1}$] }  & \multicolumn{3}{c}{ 484.8  } \\
\multicolumn{3}{l}{$K_{1}$ [km s$^{-1}$] }  & \multicolumn{3}{c}{ 31.8  } \\
\multicolumn{3}{l}{$R_\mathrm{WD}$ [\rs] }  & \multicolumn{3}{c}{ 0.01  } \\
\multicolumn{3}{l}{$log~g_{_\mathrm{WD}}$  }  & \multicolumn{3}{c}{ 8.4  } \\
\multicolumn{3}{l}{$R_\mathrm{d,out,max}$ [\rs] }  & \multicolumn{3}{c}{ 0.36  } \\
\multicolumn{3}{l}{$R_{2:1}$ [\rs]  }  & \multicolumn{3}{c}{ 0.39  } \\

\hline 

\end{tabular}
\end{table}

\begin{figure}
 \setlength{\unitlength}{1mm}
 \begin{center}
 \begin{picture}(80,140)(0,0)
 
 \put (0,124){\includegraphics[width=3.0cm,bb = 200 220 530 430, clip=]{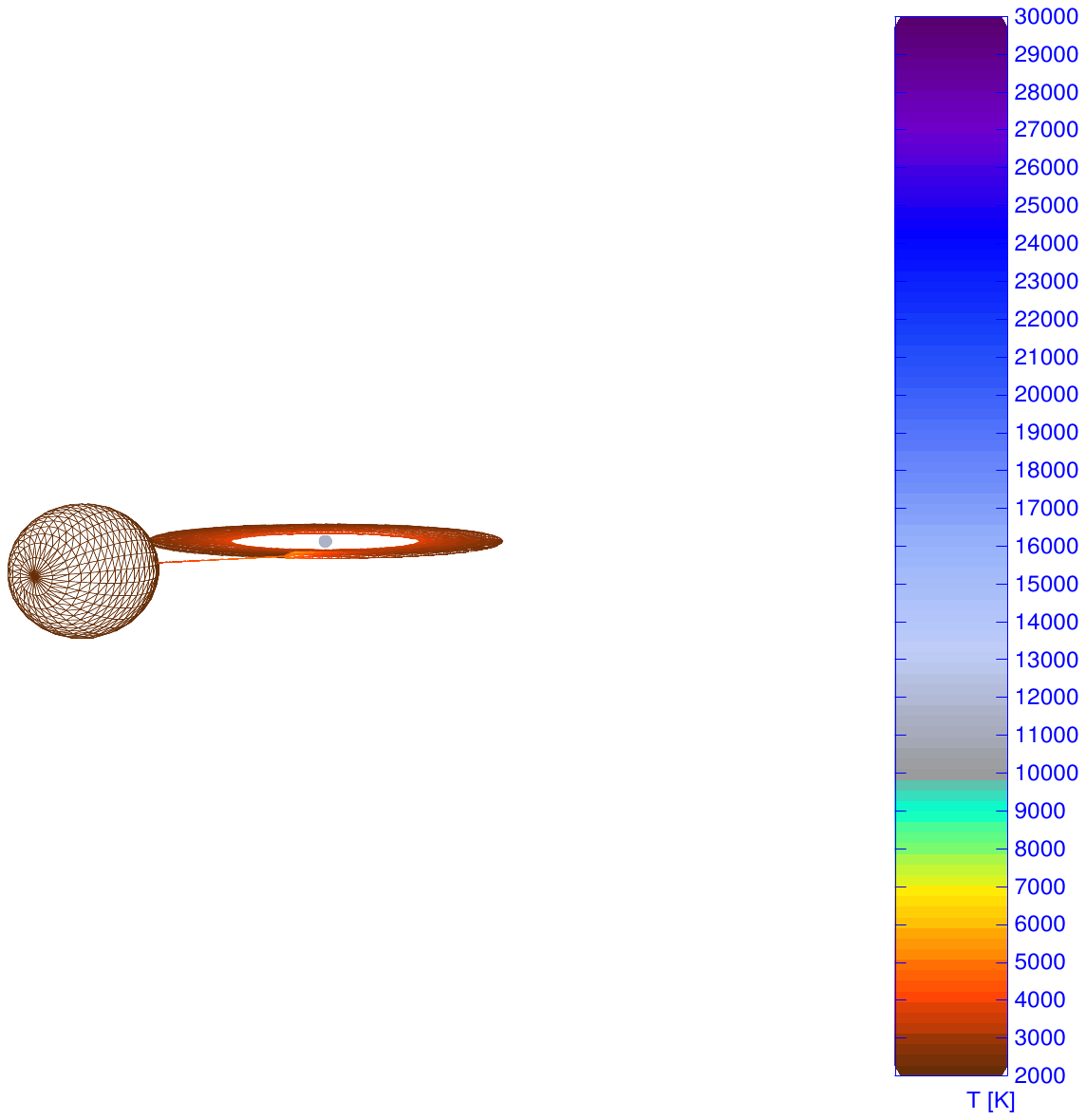}}
\put (33,124){\includegraphics[width=2.0cm,bb = 320 220 530 430, clip=]{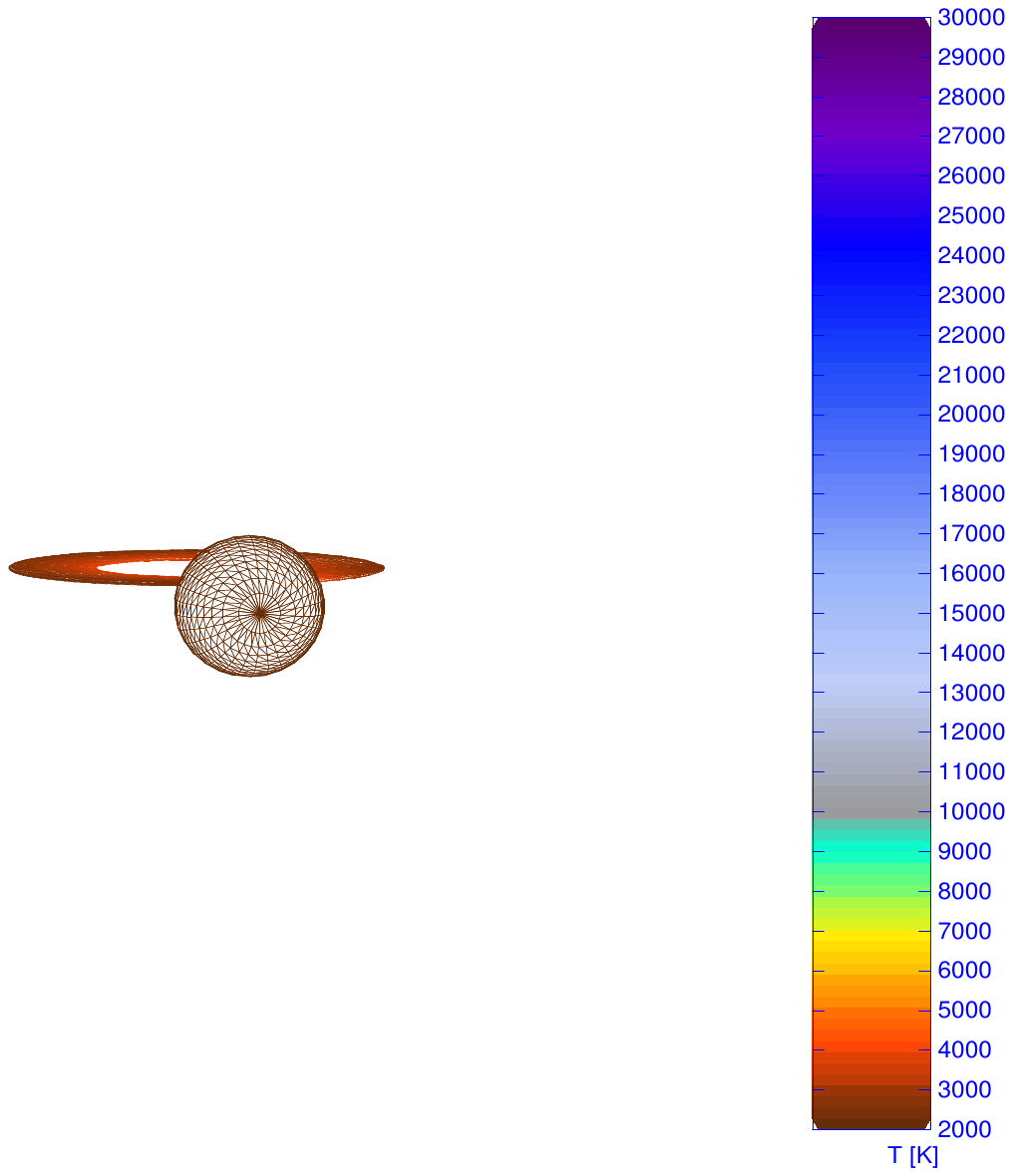}}
\put (57,124){\includegraphics[width=3.0cm,bb = 320 220 630 430, clip=]{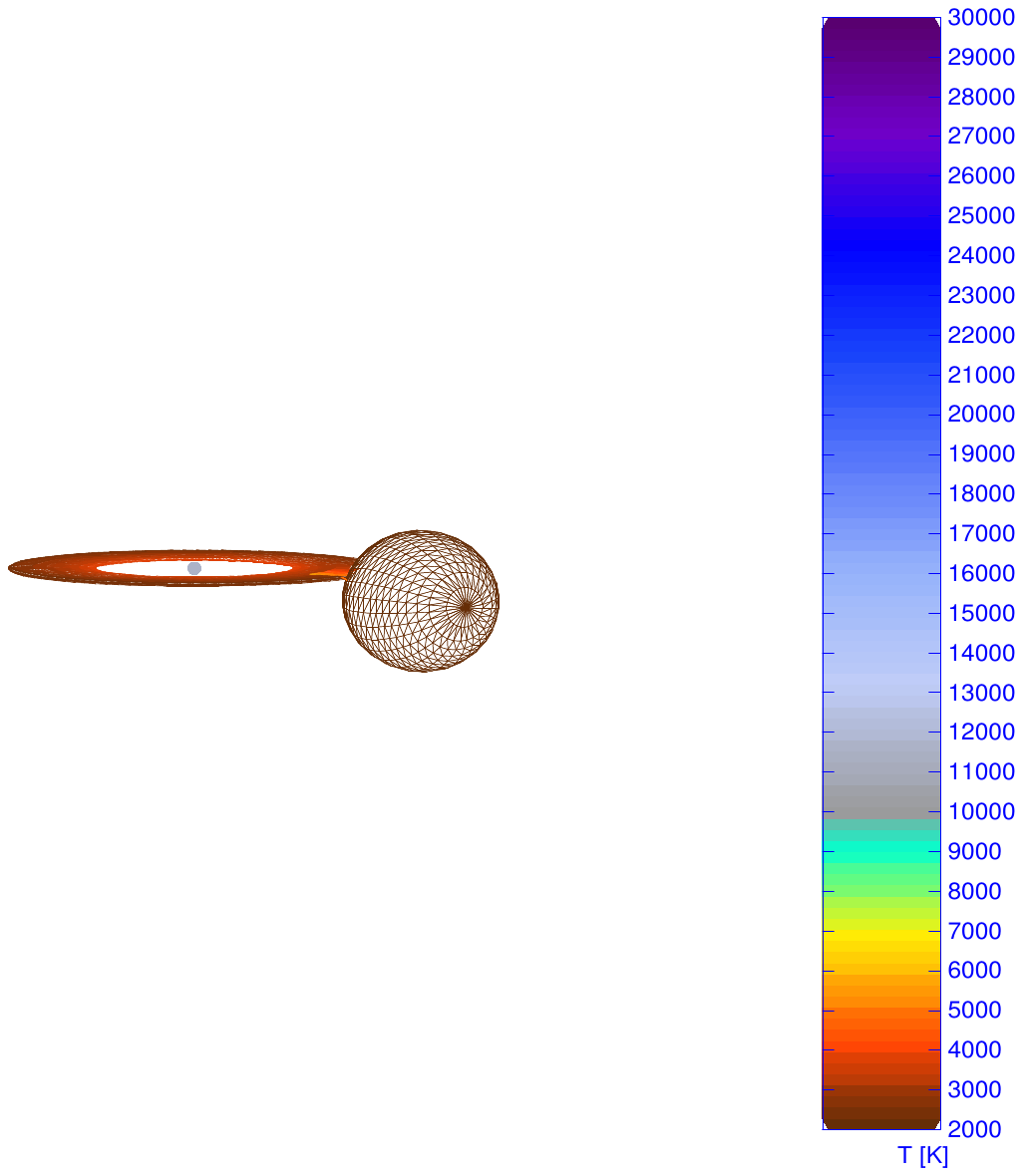}}
\put (3,63){\includegraphics[width=1.00\columnwidth,bb = 10 30 710 535, clip=]{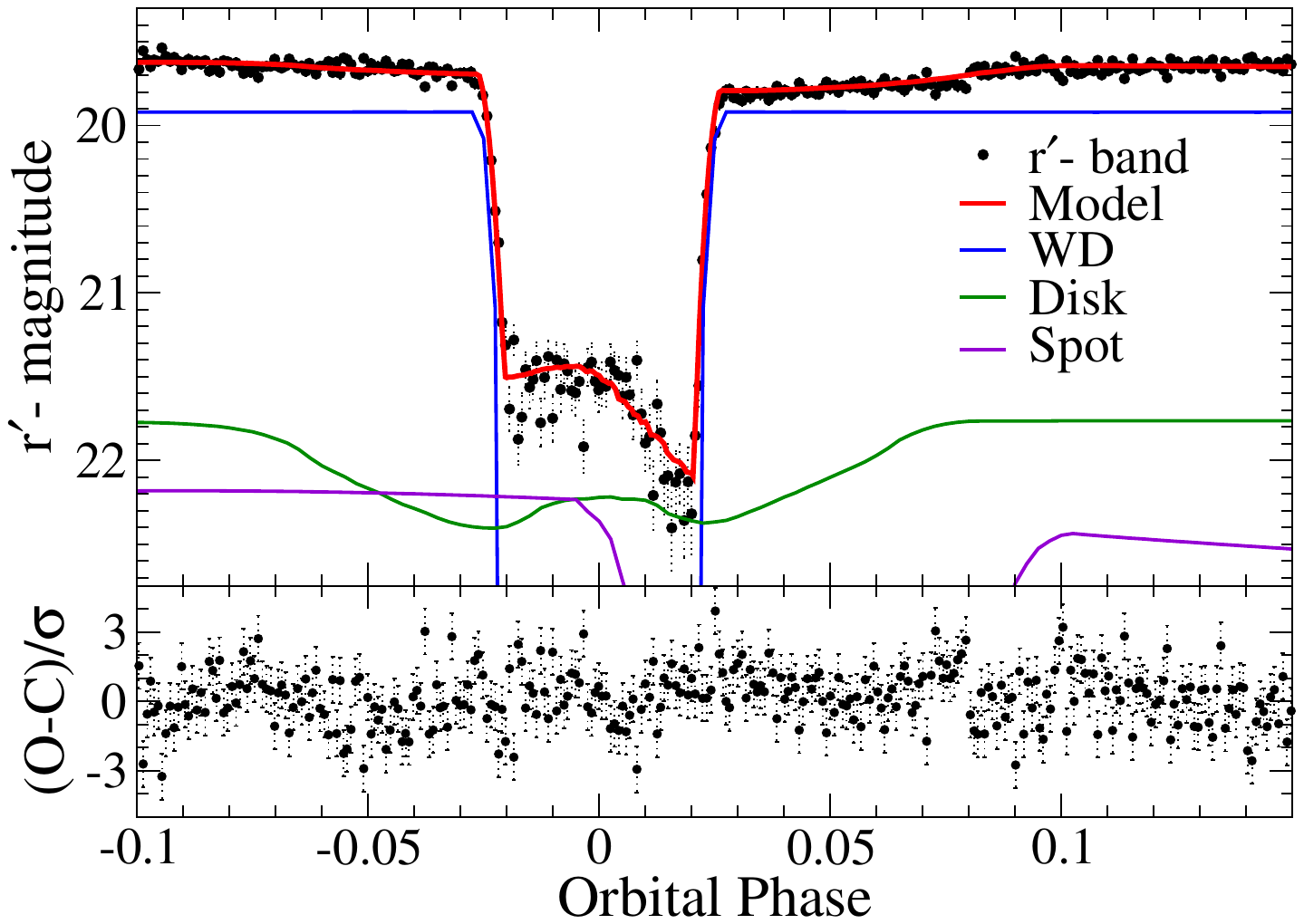}}
\put (3,0){\includegraphics[width=1.00\columnwidth,bb = 10 30 710 525, clip=]{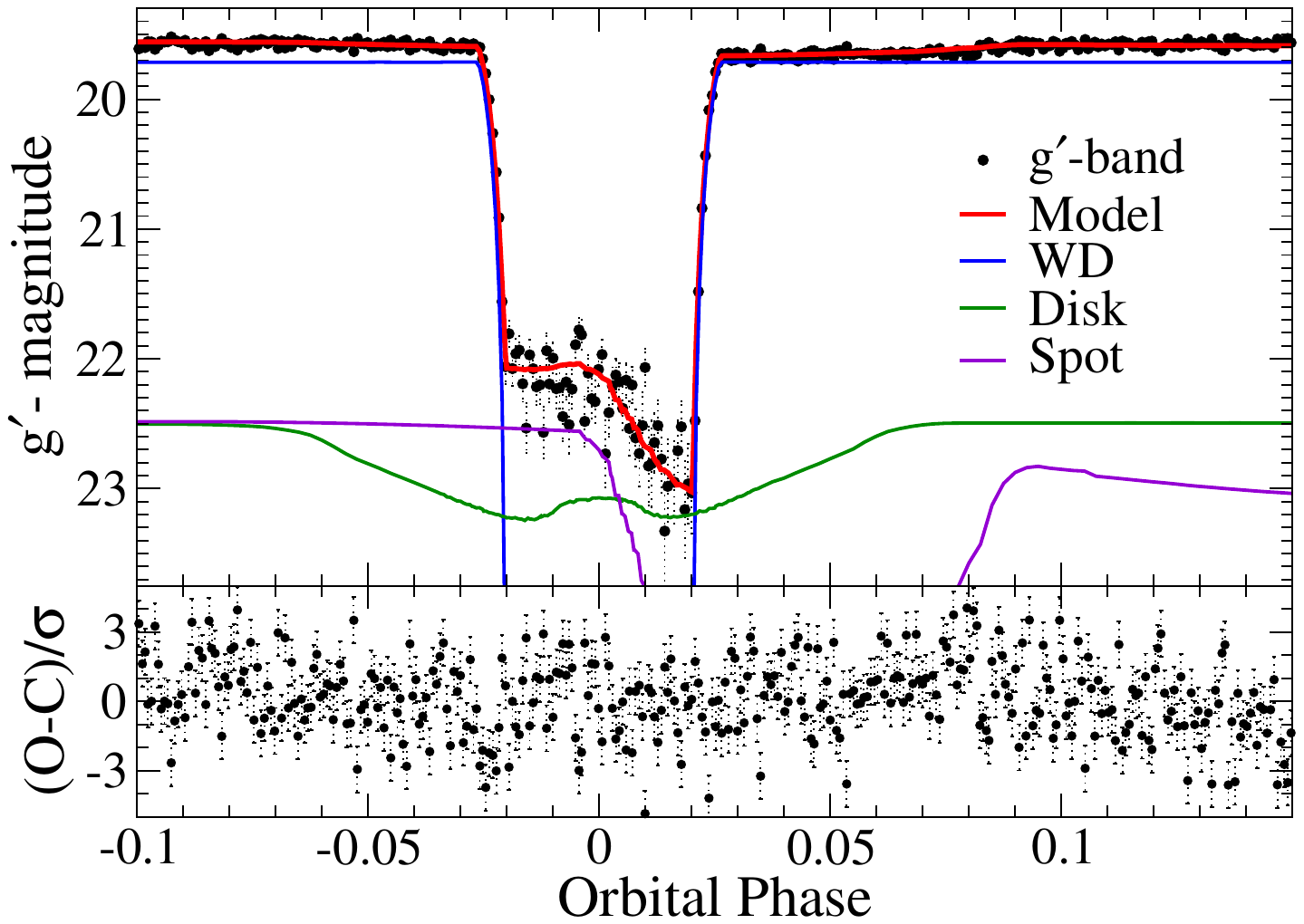}}
\put (13.5, 123) {\line(0,1){5}}
\put (47.5, 123) {\line(0,1){5}}
\put (67.5, 123) {\line(0,1){5}}
\end{picture}
\end{center}
\caption{The r$^\prime$- band and g$^\prime$-band light curves of SDSS1057 (black points) and the resulting light curves from the model (red solid lines together with (O-C)/$\sigma$  diagram of the fit. At the top appearances of the system for an observer in corresponding orbital phase are shown. }
    \label{fig:LCmod}
\end{figure}

\begin{figure}
 \setlength{\unitlength}{1mm}
 \begin{center}
 \begin{picture}(80,77)(0,0)
\put (0,0){\includegraphics[width=1.10\columnwidth,bb = 0 0 1500 1300, clip=]{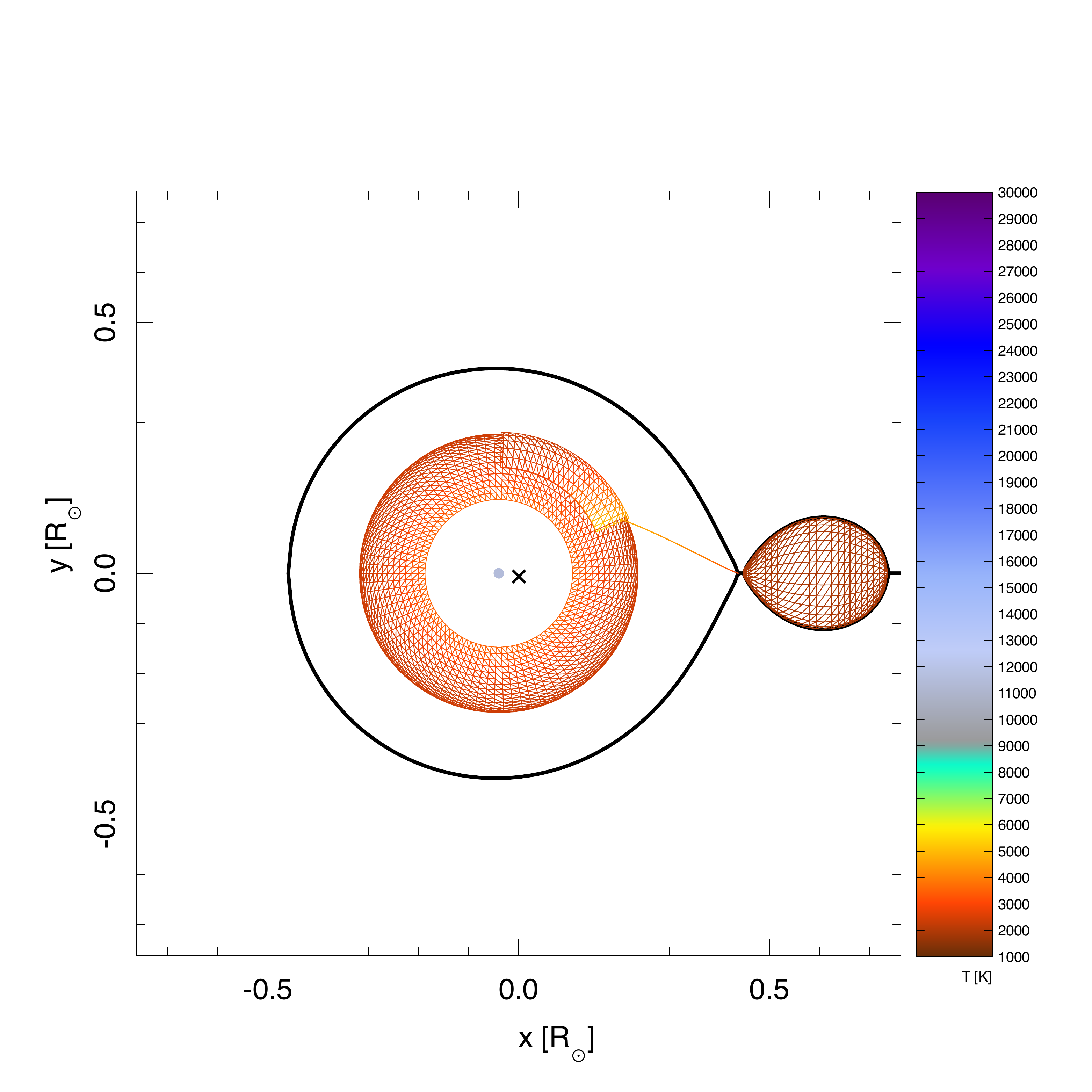}}
\end{picture}
\end{center}
\caption{The geometry of SDSS1057 system. The color show effective temperature corresponding the blackbody radiation.}
    \label{fig:LCmodGeo}
\end{figure}

\section{New Modelling}
\label{mod}

Together with our spectroscopy radial velocity results, and data obtained from the eclipses (see below), we apply a new model to \bb \,in order to try to improve our understanding of the system.

\subsection{Eclipse light-curve modeling and system parameters}
 \label{lcm}
 
The high-resolution light curves eclipses of SDSS1057 were published by \cite{mca17}. We have  obtained their data\footnote{We thank Dr. Vik Dillon and Dr. Stuart Littlefair  for kindly providing us with the original ULTRARCAM data.}  in order to apply a modeling tool developed by \citet{2013A&A...549A..77Z} and described in details in \citet{2020MNRAS.497.1475S} and \citet{2021A&A...652A..49K} to redetermine the system parameters and try to study the accretion disk structure.
The model assumes that the disk radiates as a black body at the local effective temperature with radial distribution across the disk given by the equation:
\begin{equation}
 \begin{array}{lll}
T_{\rm eff}(r) &  = &T_0 \left\{ \left(\frac{r}{R_{\rm WD}}\right)^{-3} \, \left(1-
\left[\frac{R_{\rm WD}}{r}\right]^{1/2}\right)\right\}^{EXP},  \\
T_0 & = & \left[ \frac{3GM_{\rm WD}\dot{M}}{8\pi\sigma R_{\rm WD}^3}\right]^{1/4},
\end{array}
\label{Tempeq}
\end{equation}

\noindent
where $M_{\mathrm{WD}}$ and $R_{\mathrm{WD}}$ are the mass and the radius of the WD, respectively, 
$\dot{M}$ is the mass-transfer rate, 
$G$ is the gravitational constant and 
$\sigma$ is the Stefan--Boltzmann constant. 
 In the standard accretion disk model, the radial temperature gradient is taken as $EXP$ = 0.25 \citep[equation 2.35]{1995cvs..book.....W},
but here, we allow it  to slightly deviate from this value in a manner similar to  \citet{2010ApJ...719..271L}.  
The accretion disk  thickness is defined as
\begin{equation}
z_\mathrm{d}(r) = z_\mathrm{d}(r_{\mathrm{out}})(r/r_{\mathrm{out}})^{\gamma_{\mathrm{disk}}},
\end{equation}
 
 where $\gamma_{\mathrm{disk}}$ is a free parameter for which we used the standard value of $\gamma_{\mathrm{disk}}= 9/8$ \citep[equation 2.51b]{1995cvs..book.....W} as the initial value. We also used the complex model for the hot spot described in detail in \citet[see their fig.~9 and therein]{2021A&A...652A..49K}, where all parameter definitions can be found. The disk limb-darkening used in the model follows the Eddington approximation
\citep{1980MNRAS.193..793M, 1980AcA....30..127P}.
\begin{figure}
	\includegraphics[width=1.0\columnwidth, bb = 0 0 522 530, clip=]{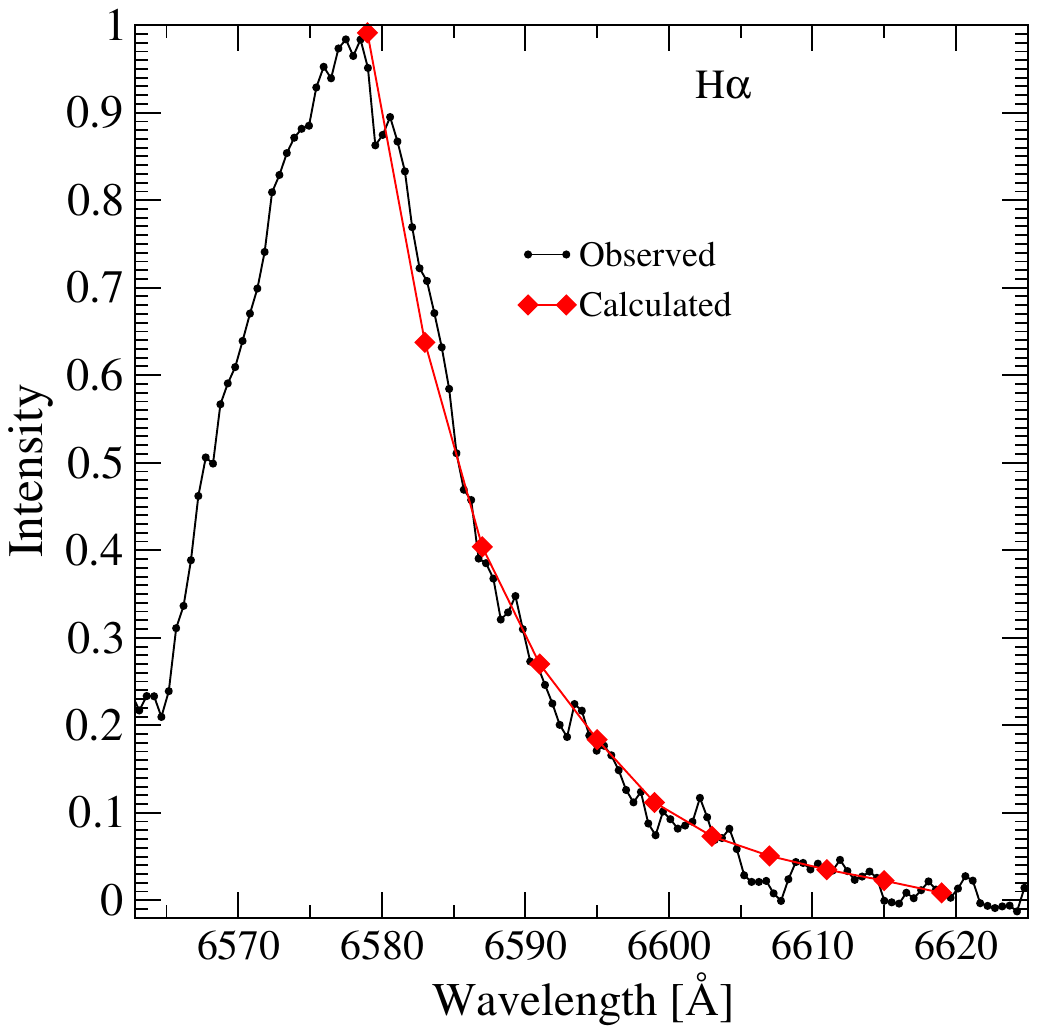}
    \caption{The red side of H$\alpha$ profile (black line) and the result of the fit of its wing by Smak's accretion disk profile (red line).}
    \label{fig:HaDiskFit}
\end{figure}

\begin{figure*}
 \setlength{\unitlength}{1mm}
 \begin{center}
 \begin{picture}(170,75)(0,0)
 \put (0,0){\includegraphics[width=\columnwidth, bb = 20 5 480 430, clip=]{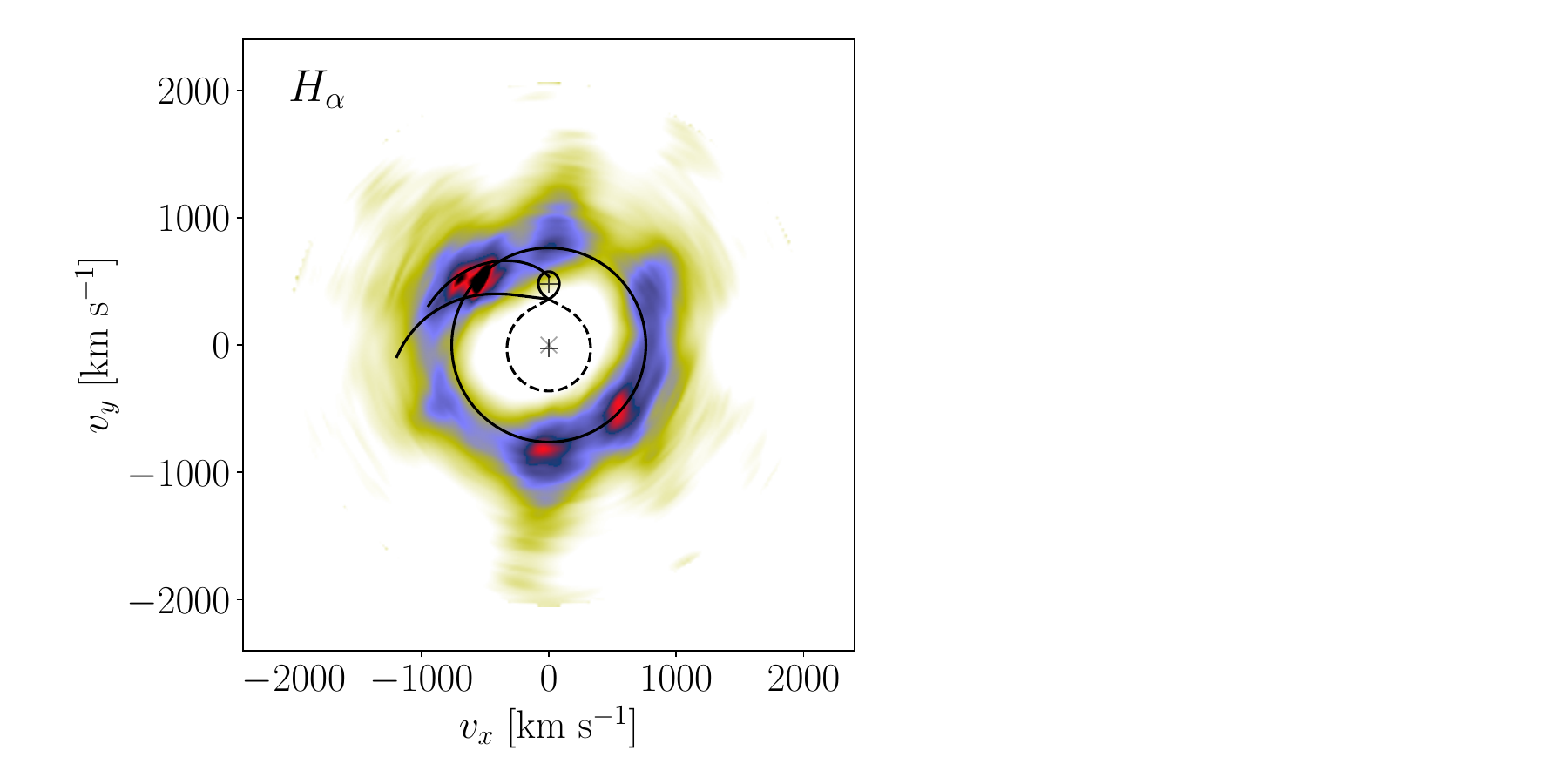}}
 \put (90,0){ \includegraphics[width=\columnwidth, bb = 20 5 480 430, clip=]{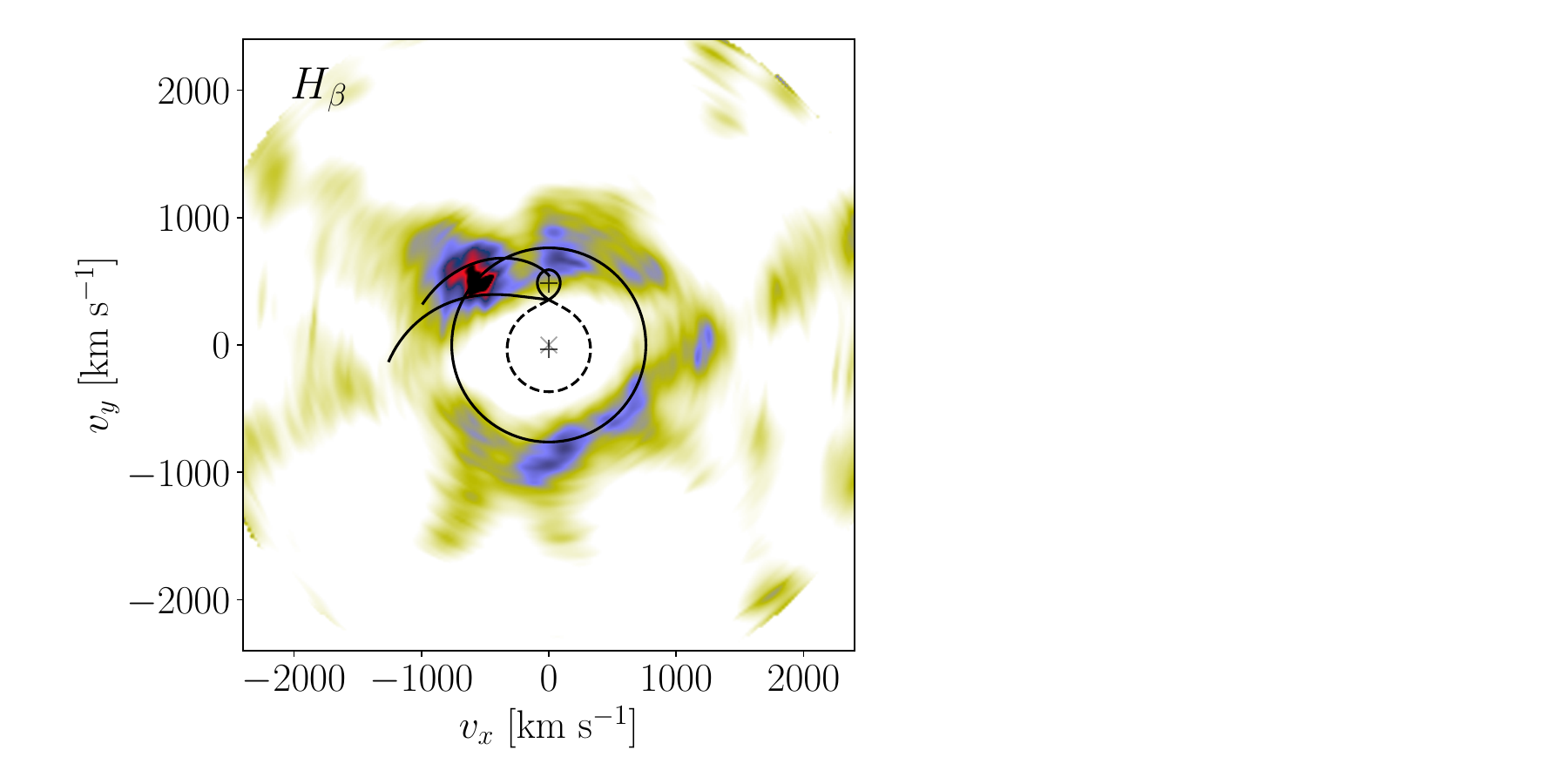}}
 \end{picture}
\end{center} 

    \caption{Doppler Tomograms of the H$\alpha$ and H$\beta$ emission lines. The relative emission intensity is shown in a scale of colors, where the strongest intensity is represented by black, followed by red, then blue, and finally yellow which depicts the weakest intensity. The color white represents the lack of emission.}
    \label{fig:tomo-hahb}
\end{figure*}

We fit independently the g$^\prime$-band  and the r$^\prime$-band light curves by searching for the minimum of the $\chi^2$ function setting all the  parameters free and using the gradient-descent method  to obtain self-consistent results for the fit.  The best parameters of the fit are given in  Table~\ref{tab:DiskPar} and the resulting light curves and contribution of each component are presented in Fig~\ref{fig:LCmod}. Numbers in brackets for the variables in Table~\ref{tab:DiskPar} are uncertainties defined as 1$\sigma$ of the Gaussian function approximation used to describe the 1-dimension $\chi^2$ function. 

The geometry of the system  is shown in Fig.~\ref{fig:LCmodGeo} for a better understanding of system/disk parameters.  The colorbar in the figure  marks the effective temperature of the blackbody, which corresponds to  radiation from the system components.
In general, our results are very close to those recently obtained by \citet{mca17}. The small differences between models are probably caused by our selection of the distance (302 pc)  compared with (367 pc) used by \citet{mca17}. We also point out that our light curve  model is in agreement with our $K_1$ estimation from the radial velocity fitting. Moreover, we note that our estimation of a mass transfer rate of 1.9$\times10^{-11}$~\ms~yr$^{-1}$ is about three times  lower than was proposed by \citet{mca17} based on the relation between WD effective temperature and a time-averaged accretion rate of \citet{2003ApJ...596L.227T, 2009ApJ...693.1007T}. Obtained result is in agreement with the expected secular mass transfer rate of $\dot{M} \sim 1.5\times10^{-11}$~\ms\ yr$^{-1}$\citep{2011ApJS..194...28K}. The main feature of the model is that a significant contribution to the continuum from the accretion disk is formed in its outer part with respect to the accretor. Another interesting feature is that the disk does  not reach the disk truncation limit $R_{d,out,max}$, and it is below  the 2:1 resonance radius $R_{2:1}$.

\begin{figure}
 \centering
 \includegraphics[width=\columnwidth, bb = 73 60 530 480, clip=]{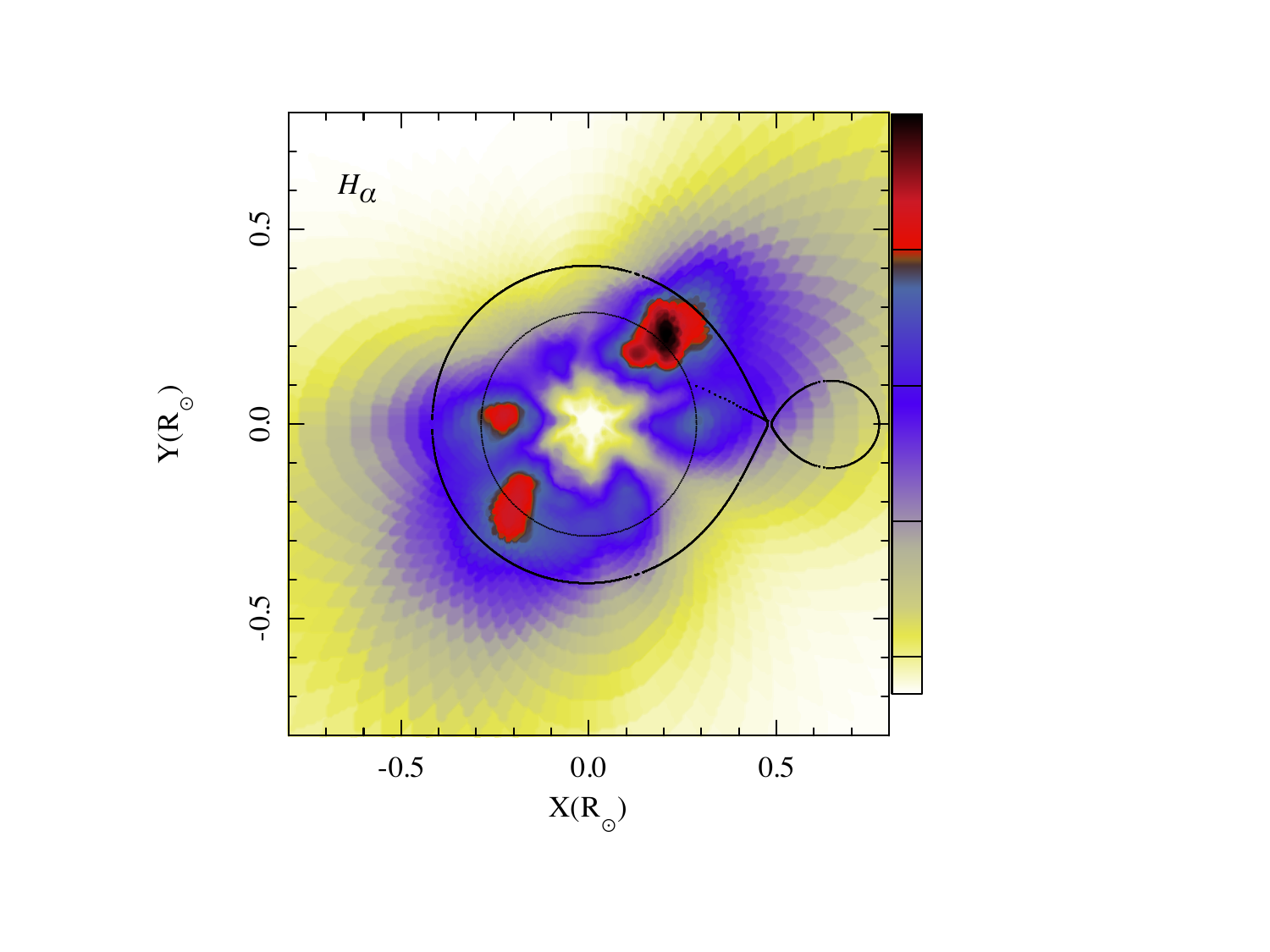}
    \caption{Brightness distribution transformed from the Doppler map of the H$\alpha$ emission line to the XY plane of the system.}
    \label{fig:Geo-ha}
\end{figure}

\subsection{Disk emission forming region}
\label{disc}

Based on the light curve fitting,  the accretion disk spectrum obtained by the removal of the underlying WD emission from the SDSS spectrum is shown in the middle panel of the Fig.~\ref{fig:HaHb}. The contribution of the disk in the r$^{\prime}$-band  out of eclipse is estimated as only  $\sim$14\%. The continuum of the accretion disk spectrum is practically flat. This continuum emission is formed, with respect of the white dwarf, in the outer part of the accretion disk with  $R_{\mathrm{d,in, continuum}}/R_{\mathrm{d,out}}=0.51$.
The effective temperature of the radiation from the disk  ranges from $\approx$3800-2400K from its inner to the outer part.
The innermost part ($\lesssim 0.5 R_\mathrm{d,out}$) of the disk is optically transparent in the continuum, similar to that observed in another bounce-back system EZ Lyn \citep{2021ApJ...918...58A}.

The Balmer decrement of the accretion disk  is found to be H$\alpha$:H$\beta$:H$\gamma$ = 1.13:1:0.95. Based on the non-LTE calculations of \citet{1980ApJS...42..351D} and \citet{1991AJ....101.1929W}, it can be demonstrated that such flat decrements can be produced in the optically thin regions of the disk with an average number density and kinetic temperature $\log$ N$_0$ $\approx$ 12.7 [cm$^{-3}$] and T$\sim$10,000 K, respectively. To estimate the emission line forming region, we used the method described by \citet{1969AcA....19..155S, 1981AcA....31..395S} and \citet{1986MNRAS.218..761H}.  In this approach, the separation between peaks in the double-peaked profiles is defined by the Keplerian velocity of the outer rim of the disk $\upsilon_{\mathrm{d,out}}$. The extent of the wings depends on the ratio of the inner to the outer radii of the disk, $R_{\mathrm{d,in}}/R_{\mathrm{d,out}}$. At the same time  its shape  is controlled by the radial emissivity profile, which is assumed to follow a  function of the form $f (r) \propto r^{-b}$, where $r$ is the radial distance from the accretor. We removed the underlying WD from the spectrum around H$\alpha$ line and fit the line wings by the above mentioned method. The steps of the fit include the calculation of Smak´s line profile; its convolution with the spectra resolution, and the minimization of the difference between  the calculated and the observed  fluxes in the wings by a discrete minimization method.  We found that the H$\alpha$ emission lines profile wings can be described by the flux from a Keplerian accretion disk with $R_{\mathrm{d,in, line}}/R_{\mathrm{d,out}} = 0.06(4)$,   $\upsilon_{\mathrm{d,out}} = 660_{-0}^{+50}$~km~s$^{-1}$, and $b=-1.37(20)$ (see Fig.~\ref{fig:HaDiskFit}).

Summarizing the result of this section, we conclude that the emission line formed in the disk fully fills the Roche lobe of the WD from practically its surface until the disk truncated radius. At the same time, the continuum emission is coming out from more of the outer part of the accretion disk.  Such a model was recently proposed by \citet{2021ApJ...918...58A} to explain the disk emission in another period bouncer EZ Lyn. 

\section{Doppler tomography}
\label{tomo}

We have constructed Doppler tomograms \citep{mah88} based on the emission line profile of the H$\alpha$ and H$\beta$ emission lines.
The maps were generated in the standard projection using code developed by \citet{1998astro.ph..6141S}. The Roche lobe of the secondary, the primary, the center mass position, and the trajectory of the stream are calculated using the system parameters obtained in previous sections. 
The results are shown in Fig.~\ref{fig:tomo-hahb}. The solid circle on the plot corresponds to the  tidally truncated radius ($R_\mathrm{d,out,max} = 0.36$\rs) of the disk which we estimated using Equation (2) from  \citet{2020A&A...642A.100N}. In both maps, there is a bright hot spot ($\upsilon_x$ = -550 km s$^{-1}$,  $\upsilon_y =$ 500 km s$^{-1}$) formed at the impact of the stream and the edge of the accretion disk. The hot spot is slightly extended inside the disk along the stream trajectory.

The H$\alpha$ map clearly shows  a nonuniform ring of disk emission extending practically to the tidal limitation radius. The H$\beta$ map looks generally similar to  H$\alpha$ except for the intensity of the disk ring. The last is probably caused by a lower signal-to-noise ratio of the data and more strongly contribution underlying Balmer absorption from the WD atmosphere. Both maps displays  excesses  at the far side of the disk, opposite the donor star.
More instance parts of excesses located at $\upsilon_x$ = -40 km s$^{-1}$,  $\upsilon_y \approx$ -790 km s$^{-1}$ and  
$\upsilon_x$ = 563 km~s$^{-1}$,  $\upsilon_y \approx -520$ km~s$^{-1}$. Similar excess, as it was mentioned by \citet{2021ApJ...918...58A}, observed in some AM CVn and period bouncer candidates.

In Fig.~\ref{fig:Geo-ha} we plot the result of transforming the $H\alpha$  Doppler map  into the system of spatial coordinates using the assumption of the Keplerian velocity distribution of the emitting particles. The color bar marks the intensity of the Doppler map. The map clearly shows emission from the disk/accretion stream impact region (the hot spot), a clumpy structure of H$\alpha$ emission from the accretion disk, and an excess of the emission at the side of the disk opposite to the hot spot. It is very interesting that the XY- map of SDSS 1057 shows likeness to the XY-map of another period bouncer EZ Lyn (see \citep[fig.14, right]{2021ApJ...918...58A}). In the case of EZ Lyn, this structure was associated with a spiral pattern in the accretion disk in which the  2:1 resonance is presented. In SDSS1057, we found that the outer part of the disk does not reach the 2:1 resonance radius ($R_{2:1} = 0.39$\rs), thus the spiral pattern can be weak and appears only in the emission lines. 
 
\section{Conclusions}
\label{sec:Concl}

We performed time-resolved spectroscopic observations of the eclipsing period bouncer SDSS J105754.25+275947.5.
Based on the obtained spectroscopy, using a new Gaia distance estimation,  and  fitting of the eclipse light curve with our model, we determined:
\begin{enumerate}

\item The system is located at a distance of 302pc and contains a white dwarf with $M_\mathrm{WD}=0.83(3) $\ms\ and surface effective temperature of 11,500(400)~K. The mass of the secondary is $M_2=0.056~$\ms\ and its effective temperature is 2100~K which corresponds to a spectral brown dwarf class of L1 or later. The system inclination is $i$~=~84\fdg3(6). Those data, obtained based on eclipse light curve modeling fit are in good agreement with our $K_1$=33(4) km~s$^{-1}$ measurement from the spectroscopy. These results are close to  recently reported by \citet{mca17}.
The mass transfer rate in the system we estimate as 1.9$\times 10^{-11}$\ms\ yr$^{-1}$.

\item Based on analyses of the SDSS and OSIRIS spectra of the object, we conclude that the optical continuum is formed predominantly by the radiation from WD. The contribution of the accretion disk is low and originates from the outer part $r \in$ [0.15~\rs, 0.29~\rs]. The Balmer emission lines are formed in a plasma of $\log$~$N_0$~ ~=~12.7~[cm$^{-3}$] and a kinetic temperature of T$\sim$~10,000~K. The size of the disk, where emission lines are formed, extends up to 0.29~\rs, a few less than the truncated limit of the disk of 0.39~\rs \, and inward the disk down to $R_\mathrm{in}\approx 2 R_\mathrm{WD}\approx 0.017$~\rs.
The trailed spectra and Doppler tomography show the {\bf presence} of the emission from the hot spot and clumpy structure of the disk with variable intensity along the disk position angle. There is an extended region at the side opposite the hot spot with two relatively bright clumps. The last is probably caused by deviation from  Keplerian motion in this part of the accretion disk.
A similar structure of the accretion disk on period bouncer EZ Lyn was recently proposed by \citet{2021ApJ...918...58A}.

Summarizing, extending the number of time-resolved spectroscopy of period bouncers, and searching for new eclipsing systems are critical to better understanding the physical condition in the accretion disk of those rare objects.
 
\end{enumerate}
\section*{Acknowledgements}
The authors are indebted to DGAPA (UNAM) for financial support, PAPIIT projects IN103120, IN113723, IN114917, and IN119323. This work is based upon observations carried out at the Observatorio Roque de los Muchachos at the Canary Islands, Spain.  

\section*{Data Availability}

The data underlying this article will be shared on reasonable request to the corresponding author.

\bibliographystyle{mnras}
\bibliography{bibliography.bib} 

\bsp
\label{lastpage}
\end{document}